\documentclass{article}
\usepackage{arxiv}
\usepackage[utf8]{inputenc} 
\usepackage[T1]{fontenc}    
\usepackage{hyperref}       
\usepackage{url}            
\usepackage{booktabs}       
\usepackage{amsfonts}       
\usepackage{nicefrac}       
\usepackage{microtype}      
\usepackage{lipsum}
\usepackage{txfonts}
\usepackage{xcolor}
\usepackage{graphicx}
\usepackage{dcolumn}
\usepackage{bm}
\usepackage{slashed}
\title{Geometric Dark Matter\thanks{Dedicated to the memory of Rahmi G{\"u}ven, a great physicist and a candid friend.}}
\author{
  Durmu{\c s} Demir\\
  Faculty of Engineering and Natural Sciences\\
  Sabanc{\i} University\\
  34956 Tuzla, {\.I}stanbul, Turkey\\
  \texttt{durmus.demir@sabanciuniv.edu}\\
\And
  Beyhan Puli{\c c}e \\
  Department of Physics\\
  {\.I}zmir Institute of Technology\\
  35430 Urla, {\.I}zmir, Turkey  \\
  \texttt{beyhanpulice@iyte.edu.tr}
  }
\begin{document}
\maketitle
\date{\today}

\begin{abstract}
The dark matter, needed for various phenomena ranging from flat rotation curves to structure formation, seems to be not only neutral and long-living but also highly secluded from the ordinary matter. Here we show that, metric-affine gravity, which involves metric tensor and affine connection as two independent fields, dynamically reduces, in its minimal form, to the usual gravity plus a massive vector field.  The vector, which interacts with only the quarks, leptons and gravity, is neutral and long-living (longer than the age of the Universe) when its mass range is $9.4\ {\rm MeV} < M_Y < 28.4\ {\rm MeV}$. Its scattering cross section from nucleons, which is some 60 orders of magnitude below the current bounds, is too small to facilitate direct detection of the dark matter. This property provides an explanation for whys and hows of dark matter searches. We show that due to its geometrical origin the $Y_\mu$ does not couple to scalars and gauge bosons. It couples only to fermions. This very feature of the $Y_\mu$ makes it fundamentally different than all the other vector dark matter candidates in the literature. The geometrical dark matter we present is minimal and self-consistent not only theoretically but also astrophysically in that its feebly interacting nature is all that is needed for its longevity. 
\end{abstract}

\section{Introduction}
Dark matter, which is roughly 5 times more than the baryonic matter \cite{Planck2018},  has been under intense theoretical \cite{dm-models} and experimental \cite{dm-search} studies since its first inference \cite{rubin}. The particle dark matter which weighs around the weak scale and which has electroweak-size couplings to the known particles (WIMP) has always been the core of the dark matter paradigm. It has been modeled in  supersymmetry \cite{susy-dm}, extra dimensions \cite{extra-dm}, and various other contexts. It has, however, revealed itself neither in direct searches \cite{dm-search} nor in collider searches \cite{coll-dm}. This negative result possibly means that dark matter falls outside the WIMP domain in that it  interacts with known matter (proton, neutron and leptons, for instance) exceedingly weakly. 

It is known that among the well-motivated candidates for vector dark matter are also hidden sector $U(1)$ gauge bosons. They have been studied in a variety of scenarios as hidden vector dark matter which interacts with the standard model fields through kinetic mixing with the photon \cite{kinetic-portal-VDM} and through Higgs portal \cite{higgs-portal-VDM}. In view of these interactions, those vector dark matter models face stringent constraints from their stability (their lifetimes must be longer than the age of the Universe and their annihilation to the standard model particles must be consistent with experimental data). 

In an attempt to understand such a dark matter scheme, we explore geometrical fields beyond the general relativity (GR). To this end, we exercise the metric-affine gravity (MAG) \cite{mag} -- an extension of GR in which the metric $g_{\mu\nu}$ and connection $\Gamma^{\lambda}_{\mu\nu}$ are independent geometrodynamical variables. One reason for this choice is that MAG is known to admit decomposition into scalars, vectors and tensors \cite{mag-decompose}. Another reason is that attempts to understand electroweak stability via gravitational completion leads to MAG \cite{mag-natural}, showing that MAG could be the gravity sector necessitated by a UV-safe quantum field theory. Our analysis shows that MAG, in its simplest ghost-free form, decomposes into GR plus a massive vector field $Y_{\mu}$,  which couples only to fermions (quarks and leptons) such that lighter the $Y_{\mu}$ smaller the couplings. This geometric vector acquires a lifetime longer than that of the Universe if its mass range is $9.4\ {\rm MeV} < M_Y < 28.4\ {\rm MeV}$  and its scattering cross section from nucleons is some 60 orders of magnitude below the current bounds \cite{dm-search}. The $Y_{\mu}$ qualifies therefore a viable dark matter candidate, well satisfying the existing bounds. 

In the recent paper \cite{Jimenez2019}, a pseudoscalar dark matter candidate is studied in MAG such that its derivative couplings to fermions arise through its couplings to the axial vector part of the torsion. The properties of the scalar depends on various model parameters due to the decomposition of the full connection. It is claimed that the coherent oscillations of the pseudoscalar can give rise to an ultra light dark matter of  mass $\approx 10^{-22}$~eV. In the present work,
we study MAG in the Palatini formalism (torsion is zero) in which decomposition of the full connection into the Levi-Civita connection plus a rank (1,2) tensor field leads to the massive vector $Y_\mu$. Our torsion-free minimal framework leads to the geometric dark matter $Y_\mu$ which depends on a single parameter. The coupling of $Y_\mu$ to fermions follows from spin connection and is universal with the same coupling parameter. The vector dark matter (geometric dark matter)  $Y_{\mu}$ in the present work is entirely different than the candidates \cite{kinetic-portal-VDM} and \cite{higgs-portal-VDM} as well as the the approach in \cite{Jimenez2019}. In particular, the geometric dark matter is not a $U(1)$ gauge boson; it stems from geometry of the spacetime. It does not couple to scalars and gauge bosons. It couples only to fermions. These features stem from its  geometrical origin, and  make it fundamentally different than the other known vector dark matter candidates. We show that due to the geometric nature of our dark matter, there is no interaction with the photon (or any other gauge boson). Therefore, we do not need to impose any selection rule (like the well-known $Z_2$ symmetry) to prevent the decay of the $Y_\mu$ into photons. Moreover, we show that the $Y_\mu$ is a geometric vector which is generated by the affine connection as a massive vector. We do not therefore need to deal with interactions due to Higgs or Stueckelberg mechanisms. It is easy to see that this keeps the present model minimal as there is no need for additional scalars which would lead to some constraints due to annihilation of vector dark matter into standard model particles through the Higgs portal or invisible decays of the standard model Higgs. 

In what follows, Sec. II explains the physical necessity of affine connection, and Sec. III builds on it by structuring the most  minimal ghost-free MAG.  Sec. IV quantizes $Y_{\mu}$ in the flat metric limit. Sec. V shows that  $Y_{\mu}$ possesses all the features required of a dark matter particle. Sec VI concludes.

\section{Necessity of Affine Connection}
The GR, whose geometry is based on the metric tensor $g_{\mu\nu}$ and its Levi-Civita connection
\begin{eqnarray}
{}^g\Gamma^{\lambda}_{\mu\nu} = \frac{1}{2} g^{\lambda\rho} \left( \partial_{\mu} g_{\nu\rho} + \partial_{\nu} g_{\rho\mu} - \partial_{\rho} g_{\mu\nu}\right),
\end{eqnarray}
is defined by the Einstein-Hilbert action
\begin{eqnarray}
S\left[g\right] = \int d^4 x\, \sqrt{-g} \frac{M_{Pl}^2}{2} g^{\mu\nu}{\mathbb{R}}_{\mu\nu}({}^g\Gamma)
\label{action-gr}
\end{eqnarray}
as a purely metrical theory of gravity. The problem  is that this action is known not to lead to the Einstein field equations. It needs be supplemented with exterior curvature \cite{ghy} because the Ricci curvature of the Levi-Civita connection 
\begin{eqnarray}
\label{ricci1}
{\mathbb{R}}_{\mu\nu}({}^g\Gamma)  =  \partial_{\lambda}{}^g\Gamma^{\lambda}_{\mu\nu} - \partial_{\nu}{}^g\Gamma^{\lambda}_{\lambda\mu} + {}^g\Gamma^{\rho}_{\rho\lambda} {}^g\Gamma^{\lambda}_{\mu\nu}  - {}^g\Gamma^{\rho}_{\nu\lambda} {}^g\Gamma^{\lambda}_{\rho\mu},
\end{eqnarray} 
obtained from the Riemann tensor as ${\mathbb{R}}_{\mu\nu}({}^g\Gamma) \equiv {\mathbb{R}}^{\lambda}_{\mu \lambda \nu}({}^g\Gamma)$,
involves second derivatives of the metric. The need to exterior curvature disrupts the action principle for GR.

The remedy,  long known to be the Palatini formalism \cite{palatini}, is to replace the Levi-Civita connection ${}^g\Gamma^{\lambda}_{\mu\nu}$ with a (symmetric) affine connection $\Gamma^{\lambda}_{\mu\nu} = \Gamma^{\lambda}_{\nu\mu}$  and restructure the Einstein-Hilbert action (\ref{action-gr}) accordingly  
\begin{eqnarray}
S\left[g, \Gamma\right] = \int d^4 x\, \sqrt{-g} \frac{M_{Pl}^2}{2} g^{\mu\nu}{\mathbb{R}}_{\mu\nu}(\Gamma)
\label{action-Palatini}
\end{eqnarray}
to find  that $\Gamma^{\lambda}_{\mu\nu}$ reduces to ${}^g\Gamma^{\lambda}_{\mu\nu}$ dynamically because $S\left[g, \Gamma\right]$ can stay stationary against variations in $\Gamma^{\lambda}_{\mu\nu}$ only if the nonmetricity vanishes, that is, only if ${}^{\Gamma}\nabla_{\lambda} g_{\mu\nu}  = 0$. This ensures that the Palatini action (\ref{action-Palatini}) is the right framework for getting the Einstein field equations. 

\section{Metric-Affine Gravity}
The Palatini formalism, a signpost showing the  way beyond the purely metrical geometry of the GR, evolves into a dynamical theory if the affine connection $\Gamma^{\lambda}_{\mu\nu}$ acquires components beyond the Levi-Civita connection. In this context, spread of $\Gamma^{\lambda}_{\mu\nu}$  into the  curvature  \cite{mag-decompose} and matter \cite{ben-bauer} sectors, for instance, leads to the MAG.  The MAG is described by the action
\begin{eqnarray}
S\left[g, \Gamma, \digamma\right] = \int d^4 x\, \sqrt{-g} \left\{\frac{M_{Pl}^2}{2} g^{\mu\nu}{\mathbb{R}}_{\mu\nu}(\Gamma) -  \frac{\xi}{4}  {\overline{\mathbb{R}}}_{\mu\nu}(\Gamma) {\overline{\mathbb{R}}}^{\mu\nu}(\Gamma) + {\mathcal{L}}\left(g, \Gamma, \digamma\right)\right\} + \Delta S
\label{action-mag}
\end{eqnarray}
in which $\mathbb{R}_{\mu\nu}(\Gamma)$ is the Ricci curvature obtained from (\ref{ricci1}) by replacing ${}^g\Gamma$  with $\Gamma$, $\mathcal{L}$ is the Lagrangian of the matter fields $\digamma$ with $\Gamma$ kinetics, and ${{\overline{\mathbb{R}}}}_{\mu\nu}(\Gamma)$ is the second Ricci curvature 
\begin{eqnarray}
\label{ricci2}
{{\overline{\mathbb{R}}}}_{\mu\nu}(\Gamma) =  \partial_{\mu}\Gamma^{\lambda}_{\lambda\nu} - \partial_{\nu}\Gamma^{\lambda}_{\lambda\mu}
\end{eqnarray} 
obtained from the Riemann tensor as ${{\overline{\mathbb{R}}}}_{\mu\nu}(\Gamma) \equiv {\mathbb{R}}^{\lambda}_{\lambda\mu \nu}(\Gamma)$. It equals the antisymmetric part of $\mathbb{R}_{\mu\nu}(\Gamma)$, and vanishes identically in the metrical geometry, ${{\overline{\mathbb{R}}}}_{\mu\nu}({}^g\Gamma) \equiv 0$. 

The $\Delta S$ in (\ref{action-mag}), containing  two-- and higher-derivative terms, has the structure 
\begin{eqnarray}
\!\!\Delta S\left[g, \Gamma\right]\! = \!\!\!\int \!\!d^4 x\, \sqrt{-g} \left\{\! A \left(g^{\mu\nu}{\mathbb{R}}_{\mu\nu}(\Gamma)\right)^2  + B {\mathbb{R}}_{\mu\nu}(\Gamma) {{\mathbb{R}}}^{\mu\nu}(\Gamma) + C {\mathbb{R}}_{\mu\nu \alpha\beta}(\Gamma) {{\mathbb{R}}}^{\mu\nu\alpha\beta}(\Gamma) + \cdots \right\}
\label{action-magp}
\end{eqnarray}
in which the leading terms, weighted by dimensionless coefficients $A$, $B$, $C$, are of similar size as the  $\xi$ term in (\ref{action-mag}). These terms, excepting $A$, are, however, dangerous in that they give ghosts in the metrical part. This is so because $\Delta S$ involves at least four derivatives of  the metric. (The $\xi$ term in (\ref{action-mag}) has no metrical contribution and remains always two-derivative.) We will hereon drop $B$, $C$ and all higher-order terms on the danger of ghosts. The $A$ term and terms containing higher powers of $g^{\mu\nu}{\mathbb{R}}_{\mu\nu}(\Gamma)$ are known to lead collectively to a scalar degree of freedom in excess of the GR \cite{fR}. In principle, there is no harm in keeping them but we drop them as they do not have any distinctive effect on the vector dark matter we shall construct. They can be included to study vector dark matter in scalar-tensor theories \cite{scalar-tensor}, and this can indeed be an interesting route. 

Now, we continue with (\ref{action-mag}) with $\Delta S$ completely dropped. The Palatini formalism implies that MAG can always be analyzed via the decomposition 
\begin{eqnarray}
\Gamma^{\lambda}_{\mu\nu} =  {}^g\Gamma^{\lambda}_{\mu\nu} + \Delta^{\lambda}_{\mu\nu}
\label{decomp}
\end{eqnarray}
where $\Delta^{\lambda}_{\mu\nu}=\Delta^{\lambda}_{\nu\mu}$ is a symmetric tensor field. Under (\ref{decomp}), the two Ricci curvatures split as
\begin{eqnarray}
\label{curv-decomp}
{\mathbb{R}}_{\mu\nu}(\Gamma) &=& {\mathbb{R}}_{\mu\nu}({}^g\Gamma) +  \nabla_{\lambda}\Delta^{\lambda}_{\mu\nu} - \nabla_{\nu}\Delta^{\lambda}_{\lambda\mu} + \Delta^{\rho}_{\rho\lambda} \Delta^{\lambda}_{\mu\nu}  - \Delta^{\rho}_{\nu\lambda} \Delta^{\lambda}_{\rho\mu},\nonumber\\
{{\overline{\mathbb{R}}}}_{\mu\nu}(\Gamma) &=&  \partial_{\mu}\Delta^{\lambda}_{\lambda\nu} - \partial_{\nu}\Delta^{\lambda}_{\lambda\mu} 
\end{eqnarray}
to put the MAG action in (\ref{action-mag}) (with $\Delta S$ dropped) into the form
\begin{eqnarray}
S\left[g, \Delta, \digamma\right] &=& \int d^4 x\, \sqrt{-g} \Big\{ \frac{M_{Pl}^2}{2} g^{\mu\nu}{\mathbb{R}}_{\mu\nu}({}^g\Gamma) - \frac{1}{4} \xi g^{\mu\alpha} g^{\nu\beta} \left(\partial_{\mu}\Delta^{\lambda}_{\lambda\nu} - \partial_{\nu}\Delta^{\lambda}_{\lambda\mu}\right)
\left(\partial_{\alpha}\Delta^{\rho}_{\rho\beta} - \partial_{\beta}\Delta^{\rho}_{\rho\alpha}\right)\nonumber\\ &+& 
\frac{M_{Pl}^2}{2} g^{\mu\nu} \left(\Delta^{\rho}_{\rho\lambda} \Delta^{\lambda}_{\mu\nu}  - \Delta^{\rho}_{\nu\lambda} \Delta^{\lambda}_{\rho\mu} \right) + {\mathcal{L}}\left(g, {}^g\Gamma, \Delta, \digamma\right)\Big\}
\label{action-mag-decomp}
\end{eqnarray}
where $\nabla_\alpha$ is the covariant derivative with respect to the Levi-Civita connection ($\nabla_{\alpha} g_{\mu\nu} = 0$). 

In the action (\ref{action-mag-decomp}), the kinetic term, proportional to $\xi$, pertains only to the vector field $\Delta^{\lambda}_{\lambda \mu}$ but the quadratic term, proportional to $M_{Pl}^2$, involves all components of $\Delta^{\lambda}_{\mu \nu}$. In fact, it shrinks to a consistent vector field theory if the quadratic term reduces to mass term of  $\Delta^{\lambda}_{\lambda \mu}$, and this happens only if 
$\Delta^{\lambda}_{\mu \nu}$ enjoys the decomposition
\begin{eqnarray}
\Delta^{\lambda}_{\mu \nu} = \frac{1}{2} \left(\Delta^{\rho}_{\rho \mu} \delta^{\lambda}_{\nu} + \delta^{\lambda}_{\mu}  \Delta^{\rho}_{\rho \nu} - 3 g^{\lambda\alpha}\Delta^{\rho}_{\rho \alpha} g_{\mu\nu}\right)
\label{delta-decomp}
\end{eqnarray}
under which the action (\ref{action-mag-decomp}) takes the form
\begin{eqnarray}
\!\!\!\!\!\!\!\!\!\!S\!\left[g, Y, \digamma\right]\! &=&\!\!\! \int \!\!d^4 x \sqrt{-g}\! \left\{\! \frac{M_{Pl}^2}{2} R(g) - \frac{1}{4} Y_{\mu\nu} Y^{\mu\nu} - \frac{3 M_{Pl}^2}{4 \xi}  Y_{\mu} Y^{\mu} - \frac{3}{2\sqrt{\xi}} \overline{f} \gamma^{\mu}\! f Y_{\mu}  + {\overline{\mathcal{L}}}\!\left(g, {}^g\Gamma, \digamma\right)\!\right\}
\label{action-mag-decomp-zprime}
\end{eqnarray}
where $R(g)\equiv g^{\mu\nu}{\mathbb{R}}_{\mu\nu}({}^g\Gamma)$ is the metrical curvature scalar, 
$Y_\mu \equiv \sqrt{\xi} \Delta^{\lambda}_{\lambda \mu}$
is a vector field generated by the affine connection, and ${\overline{\mathcal{L}}}\left(g, {}^g\Gamma, \digamma\right)$ is part of the matter Lagrangian that does not involve $Y_{\mu}$.  This action exhibits two crucial facts about the geometrical vector $Y_{\mu}$:
\begin{enumerate}
    \item First, it is obliged to be massive if gravity is to attract with the observed strength. Indeed, the Newton's constant ($G_N = (8 \pi M_{Pl}^2)^{-1}$) and the $Y_\mu$ mass  ($M_Y^2 = \frac{3}{2 \xi} M_{Pl}^2$) are both set by the Planck scale $M_{Pl}$. This action represents a rather rare case that Planck's constant sets both the gravitational scale and a particle mass. 
    
    \item Second, it couples only to fermions $f\subset \digamma$. And its couplings, originating from the spin connection through the decomposition in (\ref{delta-decomp}),  are necessarily flavor-universal. It couples to the known (leptons and quarks in the SM) and any hypothetical (say, the dark matter particle $\chi$) fermion in the same way, with the same strength. 
\end{enumerate}

\section{Quantization}
The classical setup in (\ref{action-mag-decomp-zprime}) involves two distinct fields:  The metric tensor $g_{\mu\nu}$ which leads to gravity as is the GR, and the geometrical vector $Y_\mu$ which gives rise to a fifth force that affects fermions universally. These two may well be quantized but, given the difficulties with the quantization of gravity, it would be reasonable to keep $g_{\mu\nu}$ classical yet let $Y_\mu$ be quantized. In the flat limit, for which $g_{\mu\nu}$ nears the flat metric $\eta_{\mu\nu}$, quantum field theory is full force and effect so that $Y_\mu$, along with the other fields in ${\overline{\mathcal{L}}}\left(g, {}^g\Gamma, \digamma\right)$, changes to the field operator (see, for instance, \cite{ramond})
\begin{eqnarray}
{\hat{Y}}^{\mu}(x) = \sum_{\lambda=0}^{3} \int \frac{d^3 {\vec{p}}}{(2\pi)^{3/2}} \frac{1}{\sqrt{2 \omega(\vec{p})}} \left\{{\hat{a}}(\vec{p},\lambda) \epsilon^{\mu}(\vec{p},\lambda) e^{-i p\cdot x} +  {\hat{a}}^{\dagger}(\vec{p},\lambda) \epsilon^{\mu\star}(\vec{p},\lambda) e^{i p\cdot x} \right\}
\label{Ymu-q}
\end{eqnarray}
in which the operator ${\hat{a}}(\vec{p},\lambda)$,  with the commutator  $\left[{\hat{a}}(\vec{p},\lambda), {\hat{a}}^{\dagger}(\vec{p}^{\prime},\lambda^{\prime})\right] = i \delta^{4}\left(p-p^{\prime}\right) \delta_{\lambda \lambda^{\prime}}$, annihilates a spin-1 boson of momentum $\vec{p}$, energy $\omega(\vec{p})=(M_Y^2 + \vec{p}\cdot\vec{p})^{1/2}$, polarization direction $\lambda$, and polarization sum $\sum_{\lambda=1}^{3} \epsilon^{\mu}(\vec{p},\lambda) \epsilon^{\nu\star}(\vec{p},\lambda) = \eta^{\mu\nu} - \frac{p^\mu p^\nu}{M_Y^2}$. The $Y_\mu$-quanta can be converted into or created from any fermion $f$ and its  anti-fermion $f^c$, as will be analyzed in the next section. 

\section{Geometric Dark Matter}
In this section we will study $Y_{\mu}$ to determine if it can qualify as dark matter. (The vector dark matter, as an Abelian gauge field, has been studied in \cite{Chen2014}.) To this end, the crucial factor is its lifetime.  In fact, as follows from (\ref{action-mag-decomp-zprime}) with (\ref{Ymu-q}), it decays into a fermion $f$ and its anti-particle $f^c$ with a rate
\begin{eqnarray}
\Gamma\left(Y\rightarrow f f^c\right) = \frac{N_c^f}{8\pi}\!\left(\frac{3}{2\xi}\right)^{\frac{3}{2}} 
\!\left(1 +   \frac{4\xi m_f^2}{3 M_{Pl}^2}\right) 
\!\left(1 -  \frac{8\xi m_f^2}{3 M_{Pl}^2} \right)^{\frac{1}{2}}\!  M_{Pl} 
\end{eqnarray}
where $m_f$ is the mass of the fermion and  $N_c^f$ is the number of its colors. Since $xi m_f^2 \ll M_{Pl}^2$ for all the SM fermions, this rate reduces to 
\begin{eqnarray}
\Gamma\left(Y\rightarrow f f^c\right) = \frac{N_c^f}{8\pi}  \!\left(\frac{3}{2\xi}\right)^{\frac{3}{2}}\! M_{Pl}
\end{eqnarray}
and its summation over $u$, $d$ quarks, electron, and the three neutrinos leads to the $Y_\mu$ lifetime
\begin{eqnarray}
\tau_{\scriptscriptstyle{Y}} = \frac{1}{\Gamma_{tot}}  = \frac{4 \pi}{5} \Bigg( \frac{2}{3} \Bigg)^{3/2} \frac{\xi^{3/2}}{M_{Pl}}
\end{eqnarray}
which is larger than the age of the Universe $t_{U} = 13.8 \times 10^9$ years \cite{Planck2015} if $\xi > 1.1 \times 10^{40}$. On the other side, the decay rate is prevented to go imaginary if $\xi < 1 \times 10^{41}$. These two bounds lead to the allowed mass range 
\begin{eqnarray}
9.4\ {\rm MeV} < M_Y < 28.4\ {\rm MeV}
\label{m-range}
\end{eqnarray}
across which $Y_{\mu}$ lifetime ranges from $ 4.4 \times 10^{17}~s$ to  $1.2 \times 10^{19}~s$. This means that $Y_{\mu}$ exists today to contribute to galactic dynamics and other phenomena \cite{rubin,dm-search}. Its relic density \begin{eqnarray}
\rho_{relic} = \rho_{primordial}\ e^{-\Gamma_{\rm{tot}}t_U}
\end{eqnarray}
ranges from $\rho_{primordial}/e$ (for $M_Y = 28.4\ {\rm MeV}$) to near $\rho_{primordial}$ (for $M_Y = 9.4\ {\rm MeV}$). 

The question of if $Y_\mu$ can be detected in direct searches is a crucial one. To see this, it is necessary to compute rate of scattering from nucleons. The relevant diagrams are depicted in  Fig.\ref{fig-diagrams} below.

\begin{figure}[ht!]
\centering
\includegraphics[width=.3\linewidth]{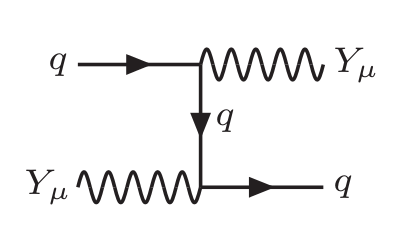} 
\includegraphics[width=.4\linewidth]{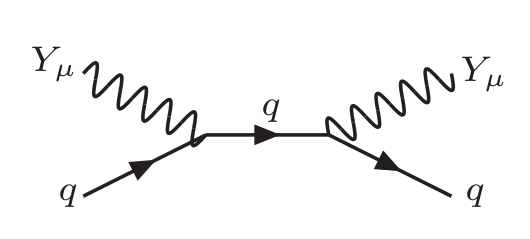}
\caption{The $Y_\mu q \rightarrow Y_\mu q$ scattering. The quark $q$ belongs to the nucleon.}
\label{fig-diagrams}
\end{figure}

\begin{figure}[ht!]
\centering
\includegraphics[width=.7\linewidth]{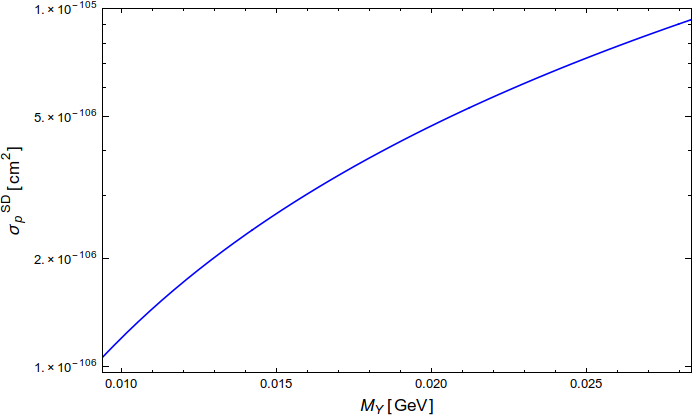} 
\caption{The spin-dependent $Y_{\mu}$--proton cross section as a function of $M_Y$.}
\label{fig-xsection}
\end{figure}

The two diagrams in Fig. \ref{fig-diagrams} result in the amplitude
\begin{eqnarray}
\label{amplitude-1}
\mathcal{M} = - i \frac{9}{4\xi} \bar{u}(k^\prime) \Bigg( \gamma^\nu \frac{\slashed{k} - \slashed{p}^\prime + m_q}{(k-p^\prime)^2-m_q^2} \gamma^\mu +  \gamma^\mu\frac{\slashed{k} + \slashed{p} + m_q}{(k+p)^2-m_q^2}\gamma^\nu \Bigg) u(k) \epsilon_\mu^*(p^\prime) \epsilon_\nu(p)
\end{eqnarray}
where $1/\xi$ in front follows from the  $Y_\mu$ coupling to quark $q$ in (\ref{action-mag-decomp-zprime}). In the nonrelativistic limit, $Y_\mu$ momenta become $p = p^\prime = (M_Y,0,0,0)$. Moreover, the quark momentum reduces to $k=(E_q,0,0,0)$ after neglecting its motion in the nucleon. The total amplitude then takes the form 
\begin{eqnarray}
\label{amplitude-2}
\mathcal{M} = - i \frac{9}{4 \xi M_Y} \bar{u}(k^\prime)(\gamma^\mu \gamma^0 \gamma^\nu - \gamma^\nu \gamma^0 \gamma^\mu ) u(k)  \epsilon_\mu^*(p^\prime) \epsilon_\nu(p)
\end{eqnarray}
after imposing $E_q \approx m_q  < M_Y$ in (\ref{amplitude-1}). As a result, the spin-dependent scattering cross section for a vector dark matter scattering off a proton \cite{Chen2014} becomes
\begin{eqnarray}
\label{spin-ind-xsection}
\sigma^{SD}_p = \frac{1}{2 \pi } \frac{m_p^2}{(M_Y + m_p)^2} a_p^2
\end{eqnarray}
where $m_p$ is the proton mass and $a_p$ is the effective spin-spin interaction of the dark matter $Y_\mu$ and the proton 
\begin{eqnarray}
a_p = \frac{9}{4 \xi M_Y} \sum_{q=u,d,s} \Delta_q^p 
\end{eqnarray}
where $\Delta_u^p = 0.84, \Delta_d^p = -0.44$ and $\Delta_s^p = -0.03$ \cite{Cheng2012}. We plot the spin-dependent cross section  (\ref{spin-ind-xsection}) in Fig.\ref{fig-xsection} in the allowed range (\ref{m-range}) of $M_Y$.   

Direct search experiments like COUPP \cite{COUPP2012}, SIMPLE \cite{SIMPLE2014}, XENON100 \cite{XENON100-2016}, PICO-2L \cite{PICO2L-2016}, PICO-60 \cite{PICO60-2017}, PandaX-II \cite{PandaX2017}, PICASSO \cite{PICASSO2017} and LUX \cite{LUX2017} have put stringent upper limits on the spin-dependent cross section for scattering of dark matter off the SM particles. They mainly exclude the WIMPs. The most stringent limit is around $\sigma_p^{SD} \sim \mathcal{O}(10^{-41})\ {\rm cm^2}$. It is clear that  the $Y_\mu$--proton spin-dependent cross section given in Fig. \ref{fig-xsection} is at most  ${\mathcal{O}}\left(10^{-106}\right)\ {\rm cm^2}$, which is too small to be measurable by any of the current experiments. The Fig. \ref{fig-xsection} can be taken as the explanation of why dark matter has so far not been detected in direct searches. It might, however, be measured in future experiments though it is hard to imagine what future technology can provide access to such tiny cross section. One possibility, speculatively speaking, would be variants of the (laser, SQUID, etc) technology that led to the detection of gravitational waves.

\section{Conclusion}
In this paper, we have set forth a new dark matter candidate, which seems to agree with all the existing bounds. In accordance with its signatures, it reveals itself only gravitationally.  Our candidate particle, a genuinely geometrical field provided by the metric-affine gravity, is a viable dark matter candidate, and explains the current conundrum by its exceedingly small scattering cross section  from nucleons. It can be difficult to detect it with today's technology but future experiments (plausibly extensions of gravitational wave detection technology) might reach the required accuracy. 

We propose a fundamentally different vector dark matter candidate from all the other vector dark matter candidates in the literature. We show that due to its geometrical origin the geometric vector dark matter $Y_\mu$ does not couple to scalars and gauge bosons. It couples only to fermions. It must be emphasized that there is no need to impose any $Z_2$ symmetry to prevent the gauge kinetic interaction in the vector portal. Its feebly interacting nature is all that is needed for its longevity. Morover, we should note that since the $Y_\mu$ is generated by the affine connection as a massive vector, we do not need to deal with interactions due to Higgs or Stueckelberg mechanisms. This keeps the present model minimal as there is no need for additional scalars conceptually. On the basis of the above-mentioned basic features it is obvious that the geometric dark matter $Y_\mu$ is the truly minimal model of dark matter. 

The model can be extended in various ways. As already mentioned in the text, one possibility is to include quadratic and higher-order terms in curvature tensor. This kind of terms, even after discarding the ghosty terms, can cause, among other things, the rank-3 tensor to be fully dynamical. The theory is then a tensor theory involving dynamical fields beyond $Y_{\mu}$, where the excess degrees of freedom may contribute to dark energy and inflation.

Before closing, it proves useful to emphasize that quantization of $Y_\mu$ is actually quantization of the geometry. But, what is done here is a partial quantization in that metric tensor is kept classical. This is certainly not the long-sought quantum gravity but it might be a glimpse of the fact that what is to be quantized may not be the metric (measurement toolbox) but the connection (the source of curvature and dark matter).
 
This work was supported in part by the T{\"U}B{\.I}TAK grant 118F387.

\end{document}